\documentclass[5p,compress]{elsarticle}

\newcommand\sect[1]{{\it #1.}---}

\usepackage[
colorlinks=true,
pdfstartview=FitV,
linkcolor=blue,
citecolor=magenta,
urlcolor=blue,
bookmarks=true,
bookmarksnumbered=true,
pdftitle={},
pdfauthor={Masaru Hongo (RIKEN)}
]{hyperref}

\usepackage{amsmath,amssymb,bm}
\usepackage{graphicx}
\usepackage{bbold}
\usepackage{color}
\usepackage[normalem]{ulem}  
\usepackage[option]{colortbl}
\usepackage{widetext}

\newcommand\diag{\operatorname{diag}}

\newcommand{\bk}{\bm{k}}
\newcommand{\bnab}{\bm{\nabla}}

\newcommand{\tile}{\tilde{e}}
\newcommand{\tilpi}{\tilde{\pi}}
\newcommand{\tils}{\tilde{S}}

\newcommand{\totAM}{J}

\newcommand{\nn}{\nonumber}
\newcommand{\zero}{ {(0)} }
\newcommand{\one}{ {(1)} }

\newcommand{\ones}{ {(1s)} }
\newcommand{\onea}{ {(1a)} }

\usepackage[normalem]{ulem}  

\renewcommand\sout{\bgroup \color{red} \ULdepth=-.5ex \ULset}

\begin{document}

\author{Koichi Hattori}\ead{koichi.hattori@outlook.com}
\address{Physics Department and Center for Field Theory and Particle Physics, Fudan University, Shanghai 200433, China}
\address{Yukawa Institute for Theoretical Physics, Kyoto University, Kyoto 606-8502, Japan}

\author{Masaru Hongo}\ead{masaru.hongo@riken.jp}
\address{RIKEN iTHEMS, RIKEN, Wako 351-0198, Japan}
\address{Research and Education Center for Natural Sciences, Keio University, Yokohama, Kanagawa 223-8521, Japan}

\author{Xu-Guang Huang}\ead{huangxuguang@fudan.edu.cn}
\address{Physics Department and Center for Field Theory and Particle Physics, Fudan University, Shanghai 200433, China}
\address{Key Laboratory of Nuclear Physics and Ion-beam Application (MOE), Fudan University, Shanghai 200433, China}

\author{Mamoru Matsuo}\ead{mamoru@ucas.ac.cn}
\address{Kavli Institute of Theoretical Sciences, University of Chinese Academy of Sciences,
19 Yuquan Road, Beijing 100049, China}
\address{RIKEN Center for Emergent Matter Science (CEMS), Wako, Saitama 351-0198, Japan}
\address{Advanced Science Research Center, Japan Atomic Energy Agency, Tokai, 319-1195, Japan}

\author{Hidetoshi Taya}\ead{h\_taya@fudan.edu.cn}
\address{Physics Department and Center for Field Theory and Particle Physics, Fudan University, Shanghai 200433, China}

\date{\today}
\title{
Fate of spin polarization in a relativistic fluid: An entropy-current analysis
}

\begin{abstract}
We derive relativistic hydrodynamic equations with a dynamical spin degree of freedom
on the basis of an entropy-current analysis.
The first and second laws of local thermodynamics
constrain possible structures of the constitutive relations including a spin current and the antisymmetric
part of the (canonical) energy-momentum tensor.
Solving the obtained hydrodynamic equations within the linear-mode analysis,
we find spin-diffusion modes, indicating that spin density is damped out after a characteristic time scale
controlled by transport coefficients introduced in the antisymmetric part of the energy-momentum tensor
in the entropy-current analysis.
This is a consequence of mutual convertibility between spin and orbital angular momentum.
\end{abstract}
\maketitle

\sect{Introduction}
Recent experimental developments opened a new avenue to study spin-dependent observables.
For example,
in relativistic heavy-ion collisions, the spin polarization of $ \Lambda $ hyperons
has been measured \cite{STAR:2017ckg}.
The result suggests that a fraction of quark spin (and perhaps gluon spin)
in the hyperons is aligned along a particular direction,
and in turn implies that a quark-gluon plasma (QGP) carries
significant magnitudes of angular momentum and/or vorticity which cause the spin alignments.
On the other hand, dynamics of spin is also a hot subject in condensed
matter physics, especially in the field of spintronics, where
generation of a spin current---a flow of spin angular momentum---is one of the key issues (see Ref.~\cite{maekawa2017spin} for a review).

To describe macroscopic dynamics of spin,
it is desirable to generalize hydrodynamics to a spinful fluid.
In the field of spintronics, a nonrelativistic framework of spin hydrodynamics has been utilized for describing the spin-current generation in the presence of a coupling between spin and vorticity
in elastic materials \cite{matsuo2013,kobayashi2017} as well as in liquid metals \cite{takahashi2016spin,matsuo2017}.
A similar framework was also used to describe the so-called micropolar fluids \cite{micropolar1,micropolar2}.
More recently, ``ideal'' relativistic hydrodynamics with spin
was proposed in the context of the QGP~\cite{Florkowski:2017ruc}
(See also Refs.~\cite{Montenegro:2017rbu,Montenegro:2017lvf,Montenegro:2018bcf} for recent attempts to apply a Lagrangian description of the spinful relativistic fluid),
in which spin is regarded as a conserved quantity in the leading-order of a gradient expansion.
However, the aforementioned nonrelativistic formulation has shown that
spin is not conserved even at the leading order.

In this Letter, we elaborate a relativistic framework
to resolve the disharmony between the nonrelativistic and relativistic formulations
and to describe the spin-dependent spacetime evolution of relativistic fluids like the QGP.
We employ a phenomenological entropy-current analysis, which was originally used to formulate relativistic viscous hydrodynamics~\cite{Landau:Fluid},
and has been recently applied to, e.g., derivation of dissipative corrections in magnetohydrodynamics (MHD)~\cite{Huang:2011dc, Hernandez:2017mch, Grozdanov:2016tdf, Hattori:2017usa}, and of the chiral magnetic/vortical effect in anomalous hydrodynamics \cite{Son:2009tf} and chiral MHD \cite{Hattori:2017usa}.
We shall show that spin density is an inherently dissipative
quantity due to mutual conversion between spin and orbital angular
momentum even in the leading-order spinful relativistic fluid.  This is a consequence of the fact that spin itself is not a conserved quantity.
It is, therefore, crucial to include dissipative terms appearing
in the first-order derivative corrections to the energy-momentum tensor
as we will discuss.
In the rest of this Letter, we first discuss the entropy-current analysis.
Then, we reinforce our observation of the spin dissipation by solving the obtained system of equations
with respect to linear perturbations applied to spin density and other hydrodynamic variables.

\sect{Phenomenological derivation of spin hydrodynamics}
Phenomenological derivation of hydrodynamics is based on the conservation laws \cite{Landau:Fluid}.
In the present case, one should consider the conservation laws of the
total angular momentum $\totAM^{\mu\alpha\beta}$
as well as the energy-momentum $\Theta^{\mu\nu} $, which are, respectively, expressed as
\begin{subequations}
\label{eq:Conservation}
\begin{align}
\label{eq:Conserv-em}
 &\partial_\mu \Theta^{\mu\nu} =0 \, ,
\\
\label{eq:Conserv-M}
&\partial_\mu \totAM^{\mu\alpha\beta} = 0 \, .
\end{align}
\end{subequations}
The total angular momentum has two contributions from orbital angular momentum and intrinsic spin.
Microscopically, they arise as two distinct components in the Noether current
for the Lorentz symmetry as
$ \totAM^{\mu\alpha\beta}
= (x^\alpha \Theta^{\mu\beta}  - x^\beta \Theta^{\mu\alpha} )
+  \Sigma^{\mu\alpha\beta} $, where $ \Sigma^{\mu\alpha\beta} = - \Sigma^{\mu\beta\alpha} $.
$ \Sigma^{\mu\alpha\beta}  $ arises from the invariance with respect to
the representation of the Lorentz group acting on a field under consideration,
and is naturally identified with an internal spin degree of freedom.
On the other hand, the orbital angular momentum in the parentheses comes
from the coordinate transformation of the argument of the field.
$ \Theta^{\mu\nu}$ in this expression is called the {\it canonical} energy-momentum tensor, which has both symmetric and antisymmetric
components: $ \Theta^{\mu\nu} \equiv \Theta^{\mu\nu}_{(s)} + \Theta^{\mu\nu}_{(a)} $.  Note that $\partial_\nu\Theta^{\mu\nu}\neq 0$ in general.

The dynamical variables near local thermal equilibrium are assumed to satisfy
the first law of thermodynamics generalized with finite spin density $ S^{\mu\nu} $~\footnote{Note that the second and third equations have slightly different physical meanings.  The differentiation $d$ in the second equation describes a transition between two different local equilibrium states, and determines physical properties of
the local thermodynamic functions (e.g., the entropy density).  On the other hand, the third equation defines the spacetime dynamics of the entropy density.
}:
\begin{eqnarray}
 &&
  Ts = e+p - \omega_{\mu\nu} S^{\mu\nu}, \quad
  T ds = de - \omega_{\mu\nu} d S^{\mu\nu},
  \nn
  \\
 &&
  T Ds = De - \omega_{\mu\nu} D S^{\mu\nu},
  \label{eq:FirstLaw}
\end{eqnarray}
where $ T $, $ s $, $e  $, and $p  $ denote
the local temperature, entropy density, energy density, and pressure, respectively.
In this Letter, we consider a neutral fluid in the absence of conserved charges.
Here, we introduced a ``spin potential'' $\omega_{\mu\nu} $
which is conjugate to the spin density $ S^{\mu\nu} $.
Note that $\omega_{\mu\nu}$ has no a priori relation to the fluid velocity $u^\mu$ at this stage
since we introduced the spin density as a new independent degree of freedom.
We also defined the derivative operator $D \equiv u^\mu \partial_\mu$,
which corresponds to the Lagrange derivative in fluid mechanics.

One may organize the constitutive relations on the basis of a derivative expansion:
\begin{subequations}
\label{eq:constitutive}
\begin{align}
&
\Theta^{\mu\nu} = e u^\mu u^\nu + p \Delta^{\mu\nu} + \Theta^{\mu\nu}_\one
\, ,
\\
&
\Sigma^{\mu\alpha\beta} =  u^\mu S^{\alpha\beta} + \Sigma^{\mu\alpha\beta} _\one
\, .
\end{align}
\end{subequations}
Here we employ the mostly plus signature of the Minkowski metric
$\eta^{\mu\nu} \equiv \diag (-1,+1,+1,+1)$.
Thus, the normalization of the fluid velocity and the transverse projection operator
read $u^\mu u_\mu = -1$ and $ \Delta^{\mu\nu} \equiv \eta^{\mu\nu} + u^\mu u^\nu $, respectively.
Here, the spin density $S^{\mu\nu}$ is assumed to satisfy the antisymmetric property $S^{\mu\nu} = - S^{\nu\mu}$
so that it has the same number of components as the total angular momentum has.
Accordingly, we have $\omega_{\mu\nu} = - \omega_{\nu\mu}$.
We introduced the first-order derivative correction to
the energy-momentum tensor $\Theta^{\mu\nu}_\one$
and the spin current $\Sigma^{\mu\alpha\beta}_\one (=-\Sigma^{\mu\beta\alpha}_\one)$.
As mentioned earlier, the former contains both symmetric and antisymmetric components,
whose roles will be elaborated below. More discussions about the tensor decomposition in Eq.~(\ref{eq:constitutive}) can be found in \ref{sec:Tensor}.

Based on the above relations, we now analyze the entropy conservation/production.
In the lowest order in the gradient, we identify the entropy current as
\begin{align}
\label{eq:s00}
s^\mu_\zero = s u^\mu
\, .
\end{align}
Therefore, by using the thermodynamic relations (\ref{eq:FirstLaw}), we find
\begin{equation}
\label{eq:s}
\partial_\mu (su^\mu) = Ds + s \theta = \beta ( De -  \omega_{\alpha\beta} D S^{\alpha\beta} ) + s \theta
\, ,
\end{equation}
where $ \theta = \partial_\mu u^\mu $. To eliminate the Lagrange derivatives,
we use equations of motion for $e$ and $S^{\alpha\beta}$ which are obtained by substituting the constitutive equations (\ref{eq:constitutive}) into the conservation laws (\ref{eq:Conservation}) and contracting Eq.~\eqref{eq:Conserv-em} with $ u_\nu $:
\begin{subequations}
\label{eq:EOMs}
\begin{align}
& \label{eq:EOMs1}
De = - (e+p) \theta + u_\nu \partial_\mu  \Theta^{\mu\nu}_\one
,
\\
& \label{eq:EOMs2}
D S^{\alpha\beta}
= -  S^{\alpha\beta} \theta
- 2  \Theta^{\alpha\beta}_\onea
-   \partial_\mu  \Sigma^{\mu\alpha\beta} _\one.
\end{align}
\end{subequations}

We use the equations of motion at the lowest-order, that is,
Eq.~(\ref{eq:EOMs1}) with $ \Theta^{\mu\nu}_\one = 0$ and Eq.~(\ref{eq:EOMs2}) with $\Sigma^{\mu\alpha\beta} _\one = 0$.
One should, however, maintain the first-order term $ \Theta^{\alpha\beta}_\onea $
in the antisymmetric part of the energy-momentum tensor
since this term is one of the lowest-order terms in Eq.~(\ref{eq:EOMs2}).
Inserting these equations into Eq.~(\ref{eq:s}),
the divergence of the entropy current can be evaluated as
\begin{align}
\label{eq:s0}
\partial_\mu s^\mu_\zero
&=
\beta \theta \{ sT - (e+p -  \omega_{\alpha\beta}  S^{\alpha\beta} )\}
+2 \beta \omega_{\alpha\beta}\Theta^{\alpha\beta}_\onea
\nn
\\
&= 2 \beta   \omega_{\alpha\beta} \Theta^{\alpha\beta}_\onea
 \, ,
\end{align}
where we used Eq.~(\ref{eq:FirstLaw}) to reach the second line.
This expression indicates that the lowest-order hydrodynamic equations of motion (\ref{eq:EOMs}) do not conserve the lowest-order entropy current (\ref{eq:s00}).  This is in contrast to the case of a fluid without spin,
where the lowest-order equations of motion correspond to ideal hydrodynamics
and describe reversible fluid dynamics.
The entropy production implies that spin density is inherently a non-conserved quantity,
which dissipates in a finite time scale.
Indeed, the conservation law of the total angular momentum (\ref{eq:Conservation})
can be cast into a ``non-conservation'' law of the spin current
\begin{equation}
\label{eq:dSigma}
\partial_\mu \Sigma^{\mu\alpha\beta} = -2 \Theta^{\alpha\beta} _{(a)}
\, .
\end{equation}
The right-hand side comes from the orbital angular momentum,
which acts as a source/absorption term for the spin current.
This is of course a natural relation indicating that
spin and orbital angular momentum are converted to each other.
The lowest-order entropy current is conserved only when $ \Theta^{\alpha\beta} _{(a)} =0 $,
i.e., when spin and orbital angular momentum are separately conserved%
\footnote{
It should be stressed that the tensors in Eq.~(\ref{eq:dSigma}) are in the canonical forms,
only in which the tensor $ \Sigma^{\mu\alpha\beta} $ corresponds to the internal spin degree of freedom.
Therefore, one {\it cannot} conclude separate conservation of the spin component, e.g.,
with a Belinfante-improved symmetric energy-momentum tensor.
}.
Since the spin current is inherently dissipative,
there is no counterpart of ideal hydrodynamics
that could be called ideal spin hydrodynamics.
In other words, spin density is not a strict hydrodynamic variable that
survives in the long time scale like a conserved charge associated with a symmetry.
Nevertheless, when the relaxation time of spin density is longer
than the mean-free-time of microscopic scattering processes,
hydrodynamic description of the transient spin dynamics is expected to work~
(see also discussions in Ref.~\cite{Becattini:2018duy}).
Such a situation should be characterized by small values of transport coefficients 
for spin dissipation, which we will introduce shortly in our formalism. 
It is, therefore, important to investigate the dissipative corrections.

For this purpose, we keep all the first-order terms in the equations of motion (\ref{eq:EOMs}). The divergence of the entropy current (\ref{eq:s}) now reads
\begin{align}
\label{eq:su1}
\hspace{-0.5cm}
\partial_\mu (su^\mu)
&=
\beta  u_\nu \partial_\mu  \Theta^{\mu\nu}_\ones
+ (\beta  u_\nu \partial_\mu  \Theta^{\mu\nu}_\onea
+ 2 \beta   \omega_{\mu\nu}  \Theta^{\mu\nu}_\onea )
\nn
\\
&\quad
+ \beta \omega_{\alpha\beta}  \partial_\mu  \Sigma^{\mu\alpha\beta} _\one
.
\end{align}
To proceed, we need to elaborate the counting scheme of the derivative expansion.
In Eq.~(\ref{eq:su1}), we regard the temperature (and the energy density) as a zeroth-order quantity.
This suggests a natural assignment
$ \omega^{\mu\nu} = {\mathcal O}(\partial^1)$
so that the two terms between the brackets fall in the same order in the gradient.
Here, we employ this counting scheme, and will accordingly drop the last term
which is an order higher as compared to the others.
Then, one can further arrange this expression as
\begin{equation}
\label{eq:s1}
\partial_\mu (s u^\mu \!+\! s^\mu_\one) \!=\!
-   \Theta^{\mu\nu}_\ones \partial_\mu \beta_\nu
\!-\!   \Theta^{\mu\nu}_\onea  (  \partial_\mu \beta_\nu \!-\! 2 \beta \omega_{\mu\nu}  ) ,
\end{equation}
where we defined $  \beta^\mu \equiv \beta u^\mu$ and identified the first-order correction to the entropy current as
$
s^\mu_\one =
-\Theta^{\mu\nu}_\one  \beta_\nu
$.
The semipositivity of this entropy production, as required by the second law of thermodynamics, can be ensured for any hydrodynamic configuration
when each term on the right-hand side has a semipositive bilinear form \cite{Landau:Fluid}.
In turn, this constrains possible tensor structures of the first-order derivative corrections
up to scalar coefficients introduced as transport coefficients.

In passing, we note that one may also employ another counting scheme with $ \omega^{\mu\nu} = {\mathcal O}(\partial^0)$.
In this counting, the spin potential $ \omega^{\mu\nu}$
modifies the constitutive relations even at the zeroth order
and provides a preferred orientation specified by $\omega^\mu=(1/2)\epsilon^{\mu\nu\rho\sigma} u_\nu\omega_{\rho\sigma}$
like a ``strong'' magnetic field $ B^\mu = {\mathcal O}(\partial^0)$ in magnetohydrodynamics~\cite{Huang:2011dc,Hernandez:2017mch,Grozdanov:2016tdf, Hattori:2017usa}.
This counting scheme suggests an interesting extension of the present work.

One may write the most general tensor structure of the first-order corrections as
\begin{subequations}
\begin{align}
&
\Theta^{\mu\nu}_\ones = 2 h^{(\mu} u^{\nu)} + \tau^{\mu\nu}
\label{eq:Theta1s}
\, ,
\\
&
\Theta^{\mu\nu}_\onea = 2 q^{[\mu} u^{\nu]} + \phi^{\mu\nu}
\label{eq:Theta1a}
\, ,
\end{align}
\end{subequations}
where $ \{  h^\mu, \tau^{\mu\nu} ,q^\mu , \phi^{\mu\nu} \} = {\mathcal O} (\partial^1) $,
$ \tau^{\mu\nu} = \tau^{\nu\mu} $, $\phi^{\mu\nu} =-\phi^{\nu\mu} $,
and $ h^\mu u_\mu = q^\mu u_\mu= \tau^{\mu\nu} u_\nu = \phi^{\mu\nu}u_\nu= 0 $.
We use the shorthand notations $ X^{( \mu\nu ) } \equiv (X^{ \mu\nu } + X^{ \nu \mu})/2   $,
$ X^{[ \mu\nu ] } \equiv (X^{ \mu\nu } - X^{ \nu \mu})/2   $,
$ X^{\langle \mu\nu \rangle}
\equiv (X^{ \mu\nu } +X^{ \nu\mu } )/2- X^\sigma_{\;\;\sigma} \Delta^{\mu\nu}/3  $.
Then, the second law of thermodynamics with spin
is guaranteed if we identify the first-order corrections as
\begin{subequations}
\label{eq:first}
\begin{align}
h^\mu &= - \kappa  ( Du^\mu + \beta \partial_\perp^\mu T)
,
\\
\tau^{\mu\nu}
  &= -  2\eta \partial_\perp^{\langle \mu} u^{\nu \rangle}   - \zeta \theta \Delta^{\mu\nu},
  \\
q^\mu &= - \lambda \big( -  D u^\mu + \beta \partial^{\mu}_\perp T
- 4 \omega^{\mu\nu} u_\nu \big)
,
\\
\phi^{\mu\nu}
  &= - 2 \gamma\big(  \partial_\perp^{[\mu} u^{\nu]}
  - 2  \Delta^\mu_\rho \Delta^\nu_\lambda \omega^{\rho\lambda}  \big)
  ,
\end{align}
\end{subequations}
with $  \partial_\perp^\mu \equiv \Delta^{\mu\nu} \partial_\nu$ and $\kappa, \eta, \zeta, \lambda, \gamma \geq 0$.
$\kappa$, $ \eta$, and $ \zeta$
are the well-known heat conductivity, shear and bulk viscous coefficients, respectively. $\lambda$ and $ \gamma$ are new transport coefficients in relativistic spin hydrodynamics.
Notice that, in the leading order of Eq.~(\ref{eq:dSigma}), the antisymmetric part $ \Theta^{\alpha\beta}_\onea $ provides
a four-dimensional torque acting on the evolution of the Lorentz generator $ \Sigma^{0\alpha\beta} $.
Therefore, the spacial projection $\phi^{\mu\nu} = \Delta^{\mu} _{\alpha} \Delta^{\nu} _{\beta}  \Theta^{\alpha\beta}_\onea $ gives rise to an antisymmetric stress which diminishes the intrinsic angular momentum of the fluid cell.
On the other hand, the temporal projection $ q^\mu =  u_{\alpha}  \Delta^\mu_{\beta} \Theta^{\alpha\beta}_\onea $
boosts the fluid cell.
For these reasons, we call $ \gamma  $ the {\it rotational viscosity}~\cite{degroot}
and $\lambda$ the {\it boost heat conductivity}.
The latter is a relativistic effect and does not have a nonrelativistic counterpart.
In global equilibrium, the entropy production should cease, so that
$ \partial^{[\mu} \beta^{\nu]} = 2 \beta \omega^{\mu\nu} $ from Eq.~(\ref{eq:s1}).  This implies that the spin potential $ \omega^{\mu\nu}$ is no longer an independent variable and is completely determined by the thermal vorticity \cite{Becattini:2012tc, Becattini:2018duy}.

Using the leading-order equation of motion for $u^\mu$, i.e., $ (e+p) Du^\mu = -  \partial^\mu_\perp p +  {\mathcal O}(\partial^2)$, we can eliminate $Du^\mu$ in $h^\mu$ and $q^\mu$ as
\begin{subequations}
\label{eq:first2}
\begin{align}
h^\mu 
 &=  - \kappa  \left[  \frac{ -   \partial^{\mu}_\perp p}{e+p} + \beta \partial^\mu_\perp T + {\mathcal O}(\partial^2)  \right]
=  {\mathcal O}(\partial^2)
,
\\
q^\mu   
 &= - \lambda  \left[  \frac{ 2  \partial^{\mu}_\perp p}{e+p}
- 4 \omega^{\mu\nu} u_\nu \right] + {\mathcal O}(\partial^2)
.
\end{align}
\end{subequations}
The heat current $h^\mu$ is beyond the first order and can be neglected within the present working accuracy
as a consequence of the spacetime translational symmetries.\footnote{
This statement is valid up to an ambiguity in the definition of $ u^\mu $ known as the ``frame choice''
(see, e.g., Refs.~\cite{Landau:Fluid, Hiscock:1985zz, Bhattacharya:2011tra, Kovtun:2012rj, Minami:2012hs}).
However, in the current case for a neutral fluid,
there are no other natural frame choices motivated by conserved currents.
}
In addition, we note that $ q^\mu $ cannot be eliminated by a frame choice at ${\mathcal O} (\partial^1)$. In fact, the variation of $ q^\mu $ under a redefinition of
the fluid velocity $ u^\mu \to u^\mu +\delta u^\mu $ with $ \delta u^\mu = {\mathcal O} (\partial^1) $ reads
$ \delta q^\mu = \delta (\Delta^\mu_{[\alpha} u_{\beta]} ) \Theta^{\alpha\beta}_{(a)} $.
Since $ \Theta^{\alpha\beta}  $ is assumed to be invariant~\cite{Bhattacharya:2011tra, Kovtun:2012rj},
this variation should be at most a second-order quantity $ \delta q^\mu = {\mathcal O} (\partial^2) $,
and is negligible as compared to $ q^\mu $ in the original frame.

\sect{Linear mode analysis}
Relativistic hydrodynamic equations with spin are obtained by plugging the constitutive relations (\ref{eq:first}) and (\ref{eq:first2})
into the conservation laws~\eqref{eq:Conservation}.
Below, we consider linear perturbations on top of global thermal equilibrium, and solve the hydrodynamic equations to discuss the dynamic evolution of the (non-)hydrodynamic modes in a spinful relativistic fluid.  Namely, we consider perturbations given by
\begin{align}
  e (x) &= e_0 + \delta e (x), \quad
  p (x) = p_0 + \delta p (x),
    \nn
  \\
  v^i (x) &= 0 + \delta v^i (x),
  \quad
  S^{\mu\nu} (x) = 0 + \delta S^{\mu\nu} (x),
  \label{eq:perturbations}
  \\
\omega^{\mu\nu} (x) &= 0 + \delta \omega^{\mu\nu} (x)
,
  \nn
\end{align}
where $v^i$ is the three-velocity of a fluid and $u^{\mu} = (1, \delta v^i) + {\mathcal O}((\delta v)^2)$.
Here, we concentrate on one of the global equilibrium configurations
where the background fluid velocity and the spin density are vanishing.
In general, finite thermal vorticity can survive in global equilibrium \cite{Becattini:2012tc}
and so is the spin potential $ \omega^{\mu\nu} $ in another counting scheme mentioned earlier.
This situation would serve as another starting point of the linear-mode analysis.

Linearizing the hydrodynamic equations
with respect to the perturbations (\ref{eq:perturbations}), we obtain
\begin{subequations}
\label{eq:linear-eqs}
\begin{align}
0 &=
  \partial_0 \delta e + \partial_i \delta \pi^i
  - 2 ( c_s^2 \lambda^\prime \partial_i \partial^i \delta e
  + D_b \partial^i \delta S^{0i} ),
  \label{eq:Energy}
  \\
0 &=
( \partial_0 \delta \pi^i + c_s^2 \partial^i \delta e)
 - \gamma_{\parallel}\partial^i \partial_j \delta \pi^j
\nn
 \\
 &\quad
- (\gamma_\perp + \gamma') (\delta^i_j \bnab^2 - \partial^i \partial_j ) \delta \pi^j
+ D_s \partial_j \delta S^{j i},
 \label{eq:Momentum}
 \\
0 &=
 \partial_0 \delta S^{ij}
 + 2 \{ D_s \delta S^{ij}
 - \gamma' (\partial^i \delta \pi^j - \partial^j \delta \pi^i ) \}
,
 \label{eq:Spin}
 \\
0 &=
  \partial_0 \delta S^{0i}
 + 2 ( c_s^2 \lambda^\prime \partial^i \delta e +D_b  \delta S^{0i} )
  ,
 \label{eq:Boost}
\end{align}
\end{subequations}
where
$\delta \pi^i \equiv \delta \Theta^{0i} = (e_0+p_0)\delta v^i
+ \lambda c_s^2 \partial^i \delta e + D_b \delta S^{0i}
$, and summation over repeated (spatial) indices are assumed.
We also introduced the constants as
\begin{align}
  &c_s^2
  \equiv
  \frac{\partial p}{\partial e}
  ,  \quad
  \chi_s    \equiv \frac{\partial S^{ij}}{   \partial \omega^{ij}  }
   , \quad
   D_s
  \equiv \frac{ 4 \gamma}{\chi_s}
 , \quad
  \gamma' \equiv \frac{\gamma}{e_0+p_0}
,
     \nn
  \\
 &
 {\chi_b}   \equiv \frac{\partial S^{i0}}{\partial \omega^{i0}  }, \quad
 {D_b}   \equiv \frac{ 4 \lambda}{{\chi_b}}, \quad
 \lambda^\prime \equiv  \frac{2\lambda }{e_0+p_0}
    ,
    \label{eq:Parameters}
  \\
&
\gamma_\parallel
 \equiv \frac{1}{e_0+p_0} \left( \zeta + \frac{4}{3} \eta \right)
 , \quad
\gamma_\perp \equiv \frac{\eta}{e_0+p_0}
  .
  \nn
\end{align}

The eigenmodes of the linearized hydrodynamic equations~\eqref{eq:Energy}-\eqref{eq:Boost}
can be obtained straightforwardly. We put the detailed calculation 
in \ref{sec:Dispersion}.
The dispersion relations of those modes read
\begin{subequations}
\label{eq:disp-relations}
\begin{align}
\label{eq:spin-diffusion}
  \omega &= - 2 iD_s
  ,
  \\
  \label{eq:boost}
\omega &= - 2 i {D_b}
,
  \\
  \omega   &=
  \begin{cases}
  \displaystyle
     - 2 iD_s - i \gamma'  k_z^2 + {\mathcal O} (k_z^4),
     \\
   - i \gamma_\perp  k_z^2 + O(k_z^4),
 \label{eq:spin-shear}
 ,
    \end{cases}
    \\
\omega &=
  \begin{cases}
\displaystyle
\pm c_s k_z - i \frac{\gamma_\parallel}{2} k_z^2 + {\mathcal O} (k_z^3),
\\
\displaystyle
- 2i {D_b} - 2i c_s^2 \lambda^\prime k_z^2 + {\mathcal O} (k_z^4).
 \end{cases}
 \label{eq:sound}
\end{align}
\end{subequations}
Note that one may take the momentum
in the $ z $-direction $\bk = (0,0,k_z)$ without loss of generality
according to the rotational symmetry of the system.
Here, we expanded the dispersion relations with respect to
the wave number $k_z$ up to ${\mathcal O}(k_z^2)$.
There are two duplicates of the second and third solutions
because of the residual rotational symmetry around $ {\bm k} $.
In total, there are ten modes, which are composed of six massive modes and four massless modes---two longitudinal and two transverse hydrodynamic modes. 
As in usual hydrodynamics, the massless modes are the shear mode and the sound mode, 
which are affected by the viscous corrections.
The gaps in the former six modes arise 
as a consequence of  the non-conservative nature of the six spin degrees of freedom $ S^{\mu\nu} $.
Therefore, we conclude that even if finite spin density is presented in a relativistic fluid,
it will be damped out after characteristic time scales $\tau_s \equiv 1/D_s$ and ${\tau_b} \equiv 1/{D_b}$.\footnote{
Note that $\tau_s$ and $\tau_b$ are different from each other 
due to the absence of the Lorentz symmetry. 
}

When $\tau_s$ and $\tau_b$ take large enough values 
compared with the typical microscopic time scale in a problem, 
spin hydrodynamics may work with the spin density being a quasi-conserved quantity. 
This means that such slow dynamics is captured by our new transport coefficients, 
the rotational viscosity $\gamma$ and boost heat conductivity $\lambda$ [cf., Eq.~(\ref{eq:Parameters})]. 
We again emphasize that there is no symmetry which guarantees separate conservation 
of the spin component out of the total angular momentum. 
The absence of a symmetry is the reason why there is no rigorous notion 
of ``ideal spin hydrodynamics'' no matter how large the spin lifetime is, 
and why possible presence of a slow spin variable depends on details of a system via the transport coefficients. 
Namely, the slowness of spin dissipation requires specific reasons 
why spin rotation in each microscopic collision process is suppressed. 
One possible reason would be suppression of spin interactions by the mass of constituent particles.

\sect{Summary and Outlook}
\label{sec:Summary}
We have derived the relativistic hydrodynamic equations
with a dynamical spin degree of freedom on the basis of
the phenomenological entropy-current analysis.
The resulting constitutive relations acquire the spin current
and the antisymmetric part of the energy-momentum tensor
as well as the usual symmetric part. We identified two new important transport coefficients
in the antisymmetric part of the energy-momentum tensor that
control the relaxation time of spin density.

We have also solved the derived spin hydrodynamic equations within the linear-mode analysis,
and found four massless hydrodynamic modes---two longitudinal propagating modes and
two transverse diffusive modes---and six non-hydrodynamic modes corresponding to the six non-conserved degrees of freedom in the spin density $S^{\mu\nu}$.

There are several interesting directions which we can pursue in future: (1) We can extend the present linear-mode analysis to the case with a finite thermal vorticity. Since such a background configuration breaks, e.g., the parity symmetry, there may appear mode mixing between (non-)hydrodynamic modes. Similarly, introducing an external or dynamical magnetic field may also lead to mode mixing or the appearance of new modes; (2) The dissipative spin hydrodynamic equations should be derived from underlying microscopic theories,
allowing for a comparison with the result presented in this Letter.
Established frameworks include the low-energy effective theory from
the local Gibbs distribution~\cite{Zubarev:1979, Becattini:2014yxa, Hayata:2015lga},
the kinetic theory based on the Wigner function formalism~\cite{Yang:2018lew, Liu:2018xip}, and the multi-moment formalism within the Boltzmann transport theory~\cite{Denicol:2018rbw} (see references therein).
c

Finally, we note that the spin hydrodynamic equations can be applied, for example,
to describe the QGP created in relativistic heavy-ion collisions and to the spintronics of
emergent relativistic quasiparticles in condensed matter physics.
For these applications, it is important to establish Kubo formulas and quantify the new transport coefficients from underlying microscopic theories, and to extend the present framework to the second order to form a causal and numerically stable system of equations.
We leave these directions as future works.

\sect{Acknowledgements}
We thank Kazuya Mameda for collaboration in the early stage of this work.
M.~H. thanks Keisuke Fujii for useful discussions. We thank Francesco Becattini, Rob Pisarski, and Dirk H. Rischke for discussions during their visits to Fudan University.
K.~H. is supported in part by China Postdoctoral Science Foundation
under Grants No.~2016M590312 and No.~2017T100266. 
M.~H. is supported by the Special Postdoctoral Researchers Program at RIKEN 
and partially by the RIKEN iTHEMS Program, in particular iTHEMS STAMP working group. 
M.~H is also supported by Japan Society of Promotion of Science (JSPS) 
Grant-in-Aid for Scientific Research (KAKENHI) Grant Numbers 18H01217, 
and the Ministry of Education, Culture, Sports, Science, and Technology (MEXT)-Supported Program for the Strategic Research Foundation at Private Universities ``Topological Science'' (Grant No. S1511006). 
X.~G.~H is supported by the Young 1000 Talents Program of China, NSFC under Grants No.~11535012 and No.~11675041.  H.~T. is partially supported by NSFC under Grants No.~11847206.

\appendix

\section{General tensor decomposition}
\label{sec:Tensor}
We here discuss the tensor decomposition in Eq.~\eqref{eq:constitutive}. Making a projection with respect to $u^\mu$, we can always write
\begin{subequations}
\label{eq:SM:decomp}
\begin{align}
\Theta^{\mu\nu} 
 &= e u^\mu u^\nu + p \Delta^{\mu\nu} + \tilde{\Theta}^{\mu\nu}
\, ,
 \\
\Sigma^{\mu\alpha\beta} 
 &=  u^\mu S^{\alpha\beta} + \tilde{\Sigma}^{\mu\alpha\beta}
\, ,
\\
s^\mu 
 &= s u^\mu + \tilde{s}^\mu
\, ,
\end{align}
\end{subequations}
in such a way that $\tilde{\Theta}^{\mu\nu}u_\mu u_\nu= \tilde{\Sigma}^{\mu\alpha\beta} u_\mu=\tilde{s}^\mu u_\mu = 0$.
Then, the corresponding divergence of the entropy current reads
\begin{equation}
\label{eq:SM:su1}
\partial_\mu s^\mu
=
\beta_\nu \partial_\mu  \tilde{\Theta}^{\mu\nu}
+ 2 \beta   \omega_{\mu\nu}  \tilde{\Theta}^{\mu\nu}
+ \beta \omega_{\alpha\beta}  \partial_\mu  \tilde{\Sigma}^{\mu\alpha\beta} + \partial_\mu \tilde{s}^\mu
.
\end{equation}
Choosing $\tilde{s}^\mu = -\beta_\nu \tilde{\Theta}^{\mu\nu}
- \beta \omega_{\alpha\beta}  \tilde{\Sigma}^{\mu\alpha\beta}$, we obtain
\begin{equation}
\label{eq:SM:su2}
\partial_\mu s^\mu
=
-\partial_\mu\beta_\nu  \tilde{\Theta}^{\mu\nu}
+ 2 \beta   \omega_{\mu\nu}  \tilde{\Theta}^{\mu\nu}
-\partial_\mu (\beta \omega_{\alpha\beta})    \tilde{\Sigma}^{\mu\alpha\beta}.
\end{equation}
Our power counting scheme assigns $\omega_{\mu\nu}={{\mathcal O}(\partial^1)}$.
Therefore, to ensure semipositive entropy production,
the transverse components must be at most $\tilde{\Theta}^{\mu\nu} \sim {{\mathcal O}(\partial^1)}$
and $\tilde{\Sigma}^{\mu\alpha\beta} \sim {{\mathcal O}(\partial^2)}$, respectively.
This shows that the tensor decomposition \eqref{eq:constitutive} does not lose generality
at ${{\mathcal O}(\partial^1)}$ within our power counting scheme,
and especially means that there are no other zeroth-order terms
which are consistent with the second law of thermodynamics.

Another issue regarding the tensor decomposition \eqref{eq:constitutive} is
that the canonical spin current $\Sigma^{\mu\alpha\beta}$ may be subject to further constraints
other than the common property $\Sigma^{\mu\alpha\beta}=-\Sigma^{\mu\beta\alpha}$.
For example, if constituent particles are Dirac fermions, one may choose $\Sigma^{\mu\alpha\beta}$ to be completely antisymmetric in all its three indices.
Nevertheless, one has the freedom to render the definition of $\Sigma^{\mu\alpha\beta}$
by using the Belinfante transformation of the energy-momentum tensor.\footnote{
We refer the readers to Refs.~\cite{Becattini:2013fla,Florkowski:2018ahw,Florkowski:2018fap} 
for more information about the meaning of the Belinfante transformation and the tensor structure of spin current. 
} 
Namely, by using an antisymmetric tensor $ G^{\lambda \mu\nu} = - G^{\mu \lambda \nu} $, 
we can define new conserved energy-momentum and total angular momentum tensors by 
\begin{subequations}
\label{eq:app:belinfante0}
\begin{align}
\Theta^{\prime\mu\nu} 
 &\equiv \Theta^{\mu\nu} + \partial_\lambda G^{\lambda \mu\nu}
\, ,
\\
\totAM^{\prime\mu\alpha\beta}
&\equiv (x^\alpha \Theta^{\prime\mu\beta}  - x^\beta \Theta^{\prime\mu\alpha} ) +  \Sigma^{\prime\mu\alpha\beta} 
\, ,
\end{align}
\end{subequations}
where the spin current reads 
\begin{equation}
 \Sigma^{\prime\mu\alpha\beta}  \equiv  \Sigma^{\mu\alpha\beta}  - 2 G^{\mu[\alpha\beta]}
 \, .
\end{equation}
Unless $ G^{\lambda \mu\nu} $ is a completely antisymmetric tensor, 
the new spin current $\Sigma^{\prime\mu\alpha\beta}   $ is 
not completely antisymmetric but is antisymmetic only in the last two indices. 
Then, one can apply the decomposition \eqref{eq:constitutive} to 
$\Theta^{\prime\mu\nu}$ and $\Sigma^{\prime\mu\alpha\beta}$ 
and perform the entropy-current analysis as in the main text. 
The choice of $ G^{\lambda \mu\nu} $ is not unique since the above transformation reduces 
the constraints on the structure of Lorentz indices in the spin current. 
However, the inverse transformation, which brings $\Sigma^{\prime\mu\alpha\beta}$ 
back to a completely antisymmetric spin current $\Sigma^{\mu\alpha\beta}$, 
requires a specific choice $ G^{\lambda \mu\nu} = \Sigma^{\prime \nu\lambda\mu}$.

\section{Dispersion relations of linear modes}
\label{sec:Dispersion}
We here summarize the linearized equations of motion \eqref{eq:linear-eqs} in a matrix form $ M \delta \vec{c} = 0$.
The Fourier components of the fluctuations are put in a vector form
$
 \delta \vec{c}
 \equiv
 (\delta \tile,
 \delta \tilpi^z,
  \delta \tils^{0z} ,
 \delta \tilpi^x,
  \delta \tils^{zx},
  \delta \tilpi^y,
  \delta \tils^{yz},
  \delta \tils^{0x} ,
  \delta \tils^{0y} ,
  \delta \tils^{xy}
  )^t
$.
The matrix $ M $ is given by a block-diagonal form
\begin{widetext}
\begin{align}
 \hspace{-0.5cm}
 M  = \left(
  \begin{array}{c|cccccccc}
A_{3\times3} &  &  &  & O  &   & &  &  \\ \hline
 &-i\omega + (\gamma_\perp + \gamma' ) k_z^2 &  + i D_s k_z &0 & 0 & 0 & 0&0 \\
 & -2  i\gamma' k_z & -i\omega + 2 D_s   & 0 & 0 &0 & 0  &0 \\
 & 0  & 0 & -i\omega + (\gamma_\perp + \gamma') k_z^2 &  - i D_s k_z & 0 & 0 & 0 \\
  O & 0 & 0 & 2 i\gamma' k_z  & -i\omega + 2  D_s & 0 &0   &0\\
  & 0 & 0 & 0 & 0 & -i\omega  + 2 {D_b} & 0&0 \\
  & 0 & 0& 0 & 0 & 0 & -i\omega  + 2 {D_b} &0\\
  & 0  & 0 & 0  &0 & 0 & 0&-i\omega + 2  D_s  \\
 \end{array}
 \right)\nn
 ,
\end{align}
\end{widetext}
where the upper left block is given by
\begin{equation}
A_{3\times3} = \left(
   \begin{array}{ccc}
  - i\omega + 2 c_s^2 \lambda^\prime k_z^2  & ik_z &   - 2 i {D_b} k_z \\
    i c_s^2 k_z & -i\omega + \gamma_\parallel k_z^2 & 0 \\
    2 i c_s^2 \lambda^\prime k_z & 0 & -i\omega  + 2 {D_b} \\
 \end{array}
 \right)
 \, .
\end{equation}
As noted in the main text, we took the momentum in the $z$-direction without loosing generality.

Thanks to the block-diagonal form of $ M $, the full secular equation
from the condition, $ {\rm det}\;M = 0 $, is immediately found to be
\begin{widetext}
\begin{equation}
( -i\omega + 2  D_s )
(-i\omega  + 2 {D_b})^2
 \left|
  \begin{array}{cc}
 -i\omega + (\gamma_\perp + \gamma') k_z^2 &  + i D_s k_z  \\
 -2  i\gamma' k_z & -i\omega + 2 D_s \\
   \end{array}
 \right|^2
\left|
  \begin{array}{ccc}
  - i\omega + 2 c_s^2 \lambda^\prime k_z^2  & ik_z &   - 2 i {D_b}k_z \\
    i c_s^2 k_z & -i\omega + \gamma_\parallel k_z^2 & 0 \\
    2 i c_s^2 \lambda^\prime k_z & 0 & -i\omega  + 2 {D_b}\\
 \end{array}
 \right|
 = 0
 \, .
\end{equation}
\end{widetext}%
The solutions of this equation provides the dispersion relations shown in Eq.~\eqref{eq:disp-relations}.
The origins of the first two factors are tracked back to
the fluctuations $ \delta \tilde S^{xy} $ and $ \delta \tilde S^{0x, 0y} $, respectively.
The third factor is from the mixing between a transverse spin and fluid velocity
$ \delta \tilde \pi^x $ and $ \delta \tilde S^{zx} $ ($ \delta \tilde \pi^y $ and $ \delta \tilde S^{zy} $).
The degeneracies in the dispersion relations occur as a consequence of
the residual rotational symmetry around the momentum $ {\bm k} $.
The last factor is from the mixing among the energy density $ \delta \tilde e $
and the longitudinal components $ \delta \tilde \pi^z $ and $ \delta \tilde S^{xy}$.

\bibliography{spin-hydro}

\end{document}